\documentclass[11pt]{article}
\usepackage{amssymb}
\usepackage{amsmath}
\usepackage{amsthm}
\usepackage{amsfonts}
\usepackage{url}
\usepackage{hyperref}
\usepackage{breakurl}

\def\GF{\mathrm{GF}}

\begin{document}
\bibliographystyle{plain}
\title{A Simple Approach to Error Reconciliation in\\ Quantum Key
Distribution%
\footnote{\hbox{Presented at the 53rd Annual Meeting of the 
Australian Mathematical Society,}
\hbox{Adelaide, 1 Oct.~2009.}  
\hbox{Copyright \copyright\ 2010 R. P. Brent.} \hspace*{\fill} rpb239}
}
\author{Richard P.\ Brent}
\date{7 May 2010}
\maketitle

\thispagestyle{empty}

\begin{abstract}
We discuss the error reconciliation phase in quantum key distribution (QKD)
and analyse a simple scheme in which blocks with bad parity (that is, blocks
containing an odd number of errors) are discarded. We predict the
performance of this scheme and show, using a simulation, that the prediction
is accurate. 
\end{abstract}

\vspace*{\fill}\pagebreak

\section{Introduction and Assumptions} \label{sec:intro}

Suppose that Alice sends $n$ random bits to Bob over a quantum channel. The
bits that Bob receives have a probability $p < 1/2$ of being 
incorrect\footnote{We do not discuss the post-selection/sifting 
phase where Alice and Bob
may discard certain bits. This requires communication over the classical
channel but relatively little computation.}. 
This could be due to noise and/or to the effect of eavesdropping by Eve.
Initially Alice and Bob have an estimate of $p$. This estimate can be
improved later, after they have some information to estimate the actual
error rate.
\smallskip

Alice and Bob want to agree on a smaller number of random bits for use as a
secret key or other cryptographic purposes. They can communicate over a
classical channel, but it is assumed that Eve can eavesdrop on all
communications over this channel (even though, in practice, it would be
protected by classical cryptography). It is assumed that communications over
the classical channel are authenticated to rule out ``man-in-the-middle''
attacks, but we do not discuss authentication here (see for
example~\cite{Taylor95,Wegman81}). Because some random bits need to be shared
between Alice and Bob for authentication purposes, QKD is more accurately 
called ``quantum key expansion''.
\smallskip

It is important that
Eve does not know the random number generator that Alice uses to generate
her $n$ random bits to send over the quantum channel~-- this random number
generator should involve some random physical device so that it is
unpredictable even if Eve has unlimited computational power.
\smallskip

Alice and Bob share a pseudo-random number generator that is used to
generate pseudo-random permutations. The seed for this random number
generator could be part of their shared initial information, or could be
sent during an earlier secure communication session. If necessary, Alice
could send Bob the key over the classical channel, {\em after} sending her
random bits over the quantum channel. Although Eve is assumed to know the
pseudo-random permutations, it is important that she can not predict them in
advance, so can not use them to decide which bits to intercept on the quantum
channel.
\smallskip

We assume that Eve is unable to store quantum states for a significant
time. Thus, any eavesdropping has to be done on the fly and can not be
delayed until Alice and Bob communicate over the classical channel.
Of course, Alice and Bob can delay communication over the classical channel
for as long as they wish, in order to make Eve's task more difficult.

\section{Expected Distribution of Errors in Blocks} 	\label{sec:dist}

Alice and Bob choose a blocksize $b$ depending on their
common estimate of $p$.  We assume $2 \le b \le n$ and for simplicity ignore
the problem of what to do with the last block if $b$ is not a divisor of $n$
(since $n$ is assumed to be large, whatever we do will make a negligible
difference to the analysis).
\smallskip

Alice and Bob apply the same random permutation to their $n$-bit
sequences, using their shared pseudo-random number generator (see above).
They should use a good random permutation 
algorithm (see Appendix~A).
\smallskip

Because of the first random permutation, we can assume that errors occurring
in a block are independent, even if the original errors are correlated.
\smallskip

We use the generating function
\[G(x) = (q + px)^b,\] 
where $q = 1-p$.
The coefficient of $x^k$ in $G(x)$ gives the probability that a block of
length $b$ contains exactly $k$ errors. Clearly this probability is
\[p^kq^{b-k}\binom{b}{k},\]
but it is convenient to avoid expressions involving sums of binomial
coefficients by working with $G(x)$.
\smallskip

Alice and Bob compute the parities of their blocks, and compare parities
using the classical channel. Thus, they can detect blocks with an odd number
of errors\footnote{Of course, Alice and Bob could use more sophisticated
error detection/correction than simple parity bits, but it is not clear that
this is desirable since it would disclose more information to Eve.}. 
We say that a block
is {\em bad} if the computed parities disagree, and {\em good} if the 
parities agree.  Note that a good block may contain an even number of
errors.
\smallskip

Let $P_0$ be the probability that a given block contains no errors.
Clearly
\[P_0 = G(0) = q^b = (1-p)^b\;.\]
Let $P_1$ be the probability that a block is bad (contains an odd number 
of errors).
Thus
\begin{equation}
P_1 = \frac{G(1) - G(-1)}{2} = \frac{1 - (1 - 2p)^b}{2} \label{eq:P1}
\end{equation}
(using $q + p = 1$, $q - p = 1 - 2p \ge 0$).
Note that, if $bp \le 1$, we have
\[P_1 = bp + O(b^2p^2)\;.\]
Let $P_2$ be the probability that a block contains errors that are not
detected (so it must contain an even number of errors).  Since
$P_0 + P_1 + P_2 = 1$, we have
\[P_2 = \frac{1 - 2(1-p)^b + (1-2p)^b}{2}
	= \frac{b(b-1)}{2}p^2 + O(b^3p^3)\;.\]
The expected number of errors in a good block is 
\[E_u = \frac{G'(1) - G'(-1)}{G(1) + G(-1)}\;,\]
where the prime indicates differentiation with respect to $x$, so
\[G'(x) = bp(q + px)^{b-1}\;.\]
Thus
\[E_u = bp\left(\frac{1 - (1 - 2p)^{b-1}}{1 + (1 - 2p)^b}\right)
	= b(b-1)p^2 + O(b^3p^3)\;.\]
Note that $E_u$ is the expected number of errors in a good block {\em before}
its first bit is discarded (see \S\ref{sec:blocksize}). 
The expected number of errors remaining after the first bit is discarded is
\[\left(\frac{b-1}{b}\right)E_u = 
   (b-1)p\left(\frac{1 - (1 - 2p)^{b-1}}{1 + (1 - 2p)^b}\right)
	= (b-1)^2p^2 + O(b^3p^3)\;.\]
After bad blocks have been discarded 
we expect the error probability for the remaining bits to be
\begin{equation}
\widetilde{p} = E_u/b = 
  p\left(\frac{1 - (1 - 2p)^{b-1}}{1 + (1 - 2p)^b}\right)
	= (b-1)p^2 + O(b^2p^3)\;. 				\label{eq:newp}
\end{equation}
The process of doing a permutation,
comparing parities and discarding some bits is called
a {\em round}. There will be several rounds, until Alice and Bob have agreed
on a string of bits that is unlikely to contain any errors\footnote{%
Actually, once Alice and Bob estimate that the expected number of
errors remaining is $\ll 1$, they will (for reasons of efficiency) adopt a
different strategy to confirm (or deny) that there are no remaining
errors~-- see~\S\ref{sec:verify}.}.

\section{Re-estimation of Error Probability}	\label{sec:reest}

Let $E_b$ be the observed block error rate, that is the number of blocks
in which an error is detected, normalised by the total number $n/b$ of
blocks\footnote{We ignore the complication that $b$ might not be a divisor
of $n$}. 
Thus the expectation ${\cal{E}}(E_b)$ 
of $E_b$ is $P_1$, and we can obtain a new
estimate $p'$ of $p$ from equation~(\ref{eq:P1}):
\[E_b = \frac{1 - (1 - 2p')^b}{2}\]
(provided $E_b < 1/2$),
which gives
\begin{equation}
p' = 
\begin{cases}
				   0 & \text{if $E_b = 0$,}\\
\left(1 - (1 - 2E_b))^{1/b}\right)/2 & \text{if $0 < E_b < 1/2$,}\\
                                 1/2 & \text{otherwise.}
\end{cases}					\label{eq:revisedp}
\end{equation}

\section{Choice of Blocksize}			\label{sec:blocksize}

In this section we consider the case that there is little or no
eavesdropping. The strategy discussed here may have to be modified if a
substantial amount of eavesdropping is detected~-- see~\S\ref{sec:block2}.
\smallskip

In our approach to error reconciliation, Alice and Bob simply discard a
block if an error is detected in it\footnote{Unlike the Cascade
algorithm~\cite[\S7]{Brassard93} (also~\cite[Ch.~3]{Sharma}), where a binary
search is performed to find an error in the block. Cascade discards fewer
correct bits, but requires more communication over the classical channel.
This is significant if the bandwidth or latency of the classical channel
is a limiting factor in the overall performance.}.
They also discard one bit, 
say the first bit, from each block in which no error is
detected, to compensate for the parity information that Eve might have
obtained about the block by eavesdropping on the classical channel. Thus,
the expected number of bits discarded per block is
\[P_1 b + (1-P_1) = 1 + P_1(b-1)\;.\]
Discarding bad blocks reduces the number of bits from $n$ to an expected
$(1 - P_1)n$. Discarding one bit from each good block reduces this further, 
to $(1 - P_1)(1 - 1/b)n$. However, to partially compensate for this reduction,
the ``quality'' of the bits should have improved.  We can quantify this in 
the following way.  
{From} Shannon's coding theorem~\cite{Shannon}
(see also \cite[\S1.2.1]{Sharma}),
the useful information
(measured in bits) contained in Bob's initial $n$ noisy bits is 
$(1 - H(p))n$, where
\begin{equation}
H(p) = -\left(p\log_2 p + q \log_2 q\right), \;\; (q = 1-p) \label{eq:entropy}
\end{equation}
is the usual Shannon entropy\footnote{We use classical Shannon entropy
throughout, although in some situations Von Neumann
entropy is appropriate~-- see~\cite[\S11.3]{Nielsen-Chuang}.},
and $p$ is the error probability.  
After discards the estimated error probability improves to $\widetilde{p}$, 
so Bob now has about
\[(1 - P_1)(1 - 1/b)(1 - H(\widetilde{p}))n\]
useful bits of information.
Dividing by $n$ to normalize, define
\begin{equation}
J(b) = (1-P_1)(1-1/b)(1-H(\widetilde{p}))\;.	\label{eq:Jb}
\end{equation}
A reasonable criterion\footnote{Strictly speaking, the coding
theorem does not apply to our situation, since Alice and Bob are trying to
agree on {\em some} common sequence of bits, and they are allowed to
exchange information over the classical channel. However, inclusion of the
entropy term in~(\ref{eq:Jb}) seems to be a useful heuristic.
See also~\cite{Maurer94}.
} for choosing $b$ is to maximise $J(b)$, subject to the
constraints that $b \ge 2$ and $b \le n$.  
If $p$ is close to $0.5$, the maximum can easily be obtained numerically
by computing $J(b)$ for $b = 2, 3, \ldots$, using equations
(\ref{eq:P1})--(\ref{eq:newp}): see Table~\ref{tab:optb}.
\begin{table}
\centering
\caption{Optimal block sizes.}                    	
\label{tab:optb}
 ~\\[0pt]
\begin{tabular}{|c|c|c|}
\hline
$p$  & $p^{-1/2}$ & $b$ \\
\hline
0.5   	& 1.41	& 2 \\
0.2	& 2.24	& 2 \\
0.1	& 3.16	& 3 \\
0.05	& 4.47	& 5 \\
0.01	& 10.0	& 10 \\
0.001	& 31.6	& 32 \\
0.0001	& 100	& 101 \\
\hline
\end{tabular}
\end{table}
\begin{table}
\centering
\caption{Crossover points for optimal  block sizes.} 	
\label{tab:crossover}
 ~\\[0pt]
\begin{tabular}{|c|c|c|}
\hline
$b$  & $p$ \\
\hline
2   	& 0.15973 \\
3	& 0.08682 \\
4	& 0.05400 \\
5	& 0.03657 \\
6	& 0.02629 \\
7	& 0.01975 \\
8 	& 0.01534 \\
9	& 0.01225 \\
10	& 0.00999 \\
\hline
\end{tabular}
\end{table}
If $bp \le 1$, then
\[J(b) = 1 + p - (bp + 1/b) + O(|bp^2\log(bp^2)|)\;,\] 
and the maximum occurs when $b \approx p^{-1/2}$. It is clear from
Table~\ref{tab:optb} that $p^{-1/2}$ is a good approximation 
for $p \le 0.1$. Table~\ref{tab:crossover} gives the crossover points
for small blocksizes $b$. The table gives, for each blocksize $b \le 10$, 
the smallest $p$ (rounded to $5$ decimals) for which that $b$ is optimal.
For example, a blocksize of 
$2$ is optimal for $0.15973 < p < 0.5$, and a blocksize of  
$9$ is optimal for $0.01225 < p < 0.01534$.
For $b$ outside the range of Table~\ref{tab:crossover}, a good approximation 
to the crossover point is $p \approx 1/b^2$.
\smallskip

Recall that the expected error probability after the first round is,
{from}~(\ref{eq:newp}),
\[\widetilde{p} = 
  p\left(\frac{1 - (1 - 2p)^{b-1}}{1 + (1 - 2p)^b}\right)\;.\]
It is interesting to consider two extreme cases.  First, suppose that $p$ is
small and $b \approx p^{-1/2}$.  Then~(\ref{eq:newp}) gives
\[\widetilde{p} = p^{3/2} + O(p^2)\;.\]
This means that the error probability converges to zero rapidly (in fact
superlinearly, with order $3/2$), provided $p$ is initially small.
\smallskip

Now consider the case that $p$ is close to $1/2$, say 
$p = 1 - q = 1/2 - \varepsilon$, where $\varepsilon$ is small but positive.
In this case we can assume that $b = 2$.
Write $\widetilde{p} = 1/2 - \widetilde{\varepsilon}$.
{From}~(\ref{eq:newp}), we have
\[\widetilde{p} = \frac{p^2}{1 - 2p + 2p^2} = \frac{p^2}{p^2 + q^2}\;,\]
which gives
\[\widetilde{\varepsilon} = \frac{2\varepsilon}{1 + 4\varepsilon^2}\;.\]
Thus, when $\varepsilon$ is small, 
$\widetilde{\varepsilon} \approx 2\varepsilon$.
After about $\log_2(1/\varepsilon)$ rounds the error probability 
will no longer be close to $1/2$.
\smallskip

Combining the analysis of the extreme cases, we see that the probability 
that any errors remain is smaller than a tolerance $\delta$ after about
\[\log_2\left(\frac{2}{1 - 2p}\right) + 
	\log_{3/2}\log_2\left(\frac{n_f}{\delta}\right)\]
rounds, where $n_f$ is the number of bits remaining after discards.
\smallskip

Table~\ref{tab:p25} gives the predicted behaviour if Alice and Bob start
with $n = 10^6$ bits, and the error probability is $p = 0.25$. The errors
are removed with five rounds, and at that point Alice and Bob share
$99642$ bits. This is before verification (described in~\S\ref{sec:verify})
and privacy amplification~(\S\ref{sec:privamp}).
\smallskip

\begin{table}
\centering
\caption{Prediction for $p = 0.25$, $n = 1000000$.} 	
\label{tab:p25}
 ~\\[0pt]
\begin{tabular}{|c|c|c|c|c|c|}
\hline
$p$  & $b$ & $n$ & errors & bad blks & new $n$ \\
\hline
0.250000 & 2  & 1000000 & 250000 & 187500 & 312500 \\
0.100000 & 3  & 312500 & 31249 	 & 25416  & 157500 \\
0.023810 & 7  & 157500 & 3749 	 & 3254   & 115470 \\
0.003532 & 17 & 115470 & 407 	 & 385    & 102507 \\
0.000201 & 71 & 102507 & 20 	 & 20	  & 99642  \\
\hline
\end{tabular}
\end{table}

To confirm the predictions made in Table~\ref{tab:p25}, we performed a
simulation.  The results of a typical run are given in
Table~\ref{tab:simulate}. 
The simulation results are in good agreement with the predictions.
\smallskip

\begin{table}
\centering
\caption{Simulation for $p = 0.25$, $n = 1000000$.} 	
\label{tab:simulate}					
 ~\\[0pt]						
\begin{tabular}{|c|c|c|c|c|c|}
\hline
$p$  & $b$ & $n$ & errors & bad blks & new $n$ \\
\hline
0.250000 & 2  & 1000000 & 250202 & 187552 & 312448 \\
0.100100 & 3  & 312448  & 31325  & 25227  & 157844 \\
0.023340 & 7  & 157844  & 3895   & 3409   & 114840 \\
0.003921 & 16 & 114840  & 406    & 386    & 101872 \\
0.000189 & 73 & 101872  & 20     & 20     & 99036  \\
\hline
\end{tabular}
\end{table}

Table \ref{tab:pvarious} shows the number of bits that we predict Alice and
Bob should agree on, for an initial block of $n = 10^6$ bits 
and various error probabilities in the range $0.0001 \le p \le 0.49$.

\begin{table}
\centering
\caption{Prediction for various $p$, $n = 1000000$.} 	
\label{tab:pvarious}
 ~\\[0pt]
\begin{tabular}{|c|c|c|c|c|}
\hline
$p$  & final $n$ \\
\hline
0.0001 	& 980197 \\
0.001  	& 928288 \\
0.01	& 761620 \\
0.10	& 318860 \\
0.20	& 152151 \\
0.25	& 99642 \\
0.30	& 56244 \\
0.35	& 33232 \\
0.40	& 14880 \\
0.45	& 3680 \\
0.48	& 587 \\
0.49	& 160 \\
\hline
\end{tabular}
\end{table}

\section{Verification}					\label{sec:verify}

After enough rounds, the estimated error probability is small, and the
expected number of remaining bit errors is less than $1$.  
At this point Alice and Bob should
verify that their bit sequences are identical. More precisely, they
should perform a probabilistic test which fails to find any discrepancy 
with extremely low probability, say $\eta$, while at the same time 
disclosing as little information as possible to Eve.
\smallskip

Alice and Bob could continue as before for about
$2\ln(1/\eta)/\ln(n)$ further rounds 
(where $n$ is the number of bits remaining),
but this would be very inefficient and would unnecessarily
disclose many parity bits (that is, linear\footnote{Parity information
is a linear relation over the field $\GF(2)$. If Eve gets enough 
such relations, she can solve for the unknown bits using linear algebra
over $\GF(2)$.} relations between the bits) to
Eve, who is assumed to be eavesdropping on the classical channel. It is much
better for Alice and Bob to compute a suitable hash of their data and then
compare this hash.  If a good $64$-bit hash agrees, then the probability
that any undetected discrepancies remain should be 
of order $2^{-64} \approx 5 \times 10^{-20}$.
\smallskip

One possibility for a $k$-bit hash function is to compute the parities of
$k$ randomly chosen subsets (each of size about $n/2$, where $n$ is the
number of bits to be verified). Each bit of the hash
can be computed efficiently by generating a pseudo-random sequence of
$n$ bits, performing a bitwise ``and'' with the data, and computing the
parity of the result\footnote{For the sake of efficiency, 
the logical operations should be performed using full-word operations.}.
\smallskip

Random-subset hashing is inefficient because only one bit of the hash is 
generated for each pass through the data. Alternatives exist that are 
about as good and much faster in practice~\cite{rpb147,Rabin,wiki2}.
\smallskip

If the verification phase fails to confirm that Alice and
Bob have identical sequences of bits, it is necessary to return to 
computing parities of blocks (of size $b \le n^{1/2}$) to eliminate the
remaining error(s), then try verification~again.

The number of bits communicated over the classical channel during the
verification phase(s) should be taken into account when estimating the 
information available to Eve. See the remarks at the 
end of~\S\ref{sec:eveinfo}.
\smallskip

\section{Summary}					\label{sec:summary}

In the following summary, all communication between Alice and Bob is over 
the classical
channel except for step~\ref{step:begin}, which uses the quantum channel. 
It is assumed that Eve can eavesdrop on
the classical channel. ``Both'' means both Alice and Bob, performing 
identical steps using the same algorithm, and obtaining the same results
(except for the block parities computed at step~\ref{step:parities}).
For example, it is crucial that Alice and Bob use the same blocksizes and 
the same random permutations.

\begin{enumerate}
\item \label{step:begin} 
	Alice sends Bob $n$ bits (where $n$ is a predetermined number) 
	over the quantum channel.
\item	Optionally, the following steps can be delayed for as long
	as Alice and Bob wish (see the remark at the end 
	of~\S\ref{sec:intro}).
\item 	Both set the estimated error probability $p$ to a predetermined
	constant.
\item \label{step:seed}	
	Both initialise their pseudo-random number generator
	with the same seed (either part of their initially shared
	information, or communicated on the classical channel after 
	step~\ref{step:begin}).
\item \label{step:loop}
	If $n$ is too small, the process fails (as in step~\ref{step:end}).  
	Otherwise, both apply a pseudo-random permutation to their $n$
	bits, as described in~\S\ref{sec:dist}.
\item 	Both compute the optimal block size $b$ as described in
	\S\ref{sec:dist}, subject to $2 \le b \le n^{1/2}$.
	If necessary, the last block is padded with zeros which will be
	removed at step~\ref{step:delete}. (See also~\S\ref{sec:eveinfo} for
	the choice of blocksize.)
\item \label{step:parities}	
	Both compute parities of their blocks and exchange these parities.
	Both then compare parities and identify bad blocks (that is, blocks 
	with an odd number of errors). 
\item \label{step:delete}
	Both delete zero padding from the last block if it is a good block.
	Both delete the bad blocks and also delete the first bit of each 
	good block. 
	Let $\widehat{n}$ be the number of bits remaining.
\item 	Both compute a new estimate $p'$ using 
	equation~(\ref{eq:revisedp}) and the observed block error rate $E_b$
	(the number of bad blocks divided by the 
	total number $\lceil n/b \rceil$ of blocks).
 	Both set $p \leftarrow p'$, and $n \leftarrow \widehat{n}$. 
\item 	Both compute an estimated error probability $\widetilde{p}$ 
	for the remaining bits, using equation~(\ref{eq:newp}).
 	Both set $p \leftarrow \widetilde{p}$.
 	Both return to step \ref{step:loop} if $p \ge 1/n$, 
	otherwise they continue with step \ref{step:verify}.
\item \label{step:verify}
	Both perform verification as described in~\S\ref{sec:verify}.
      	If verification fails, both set $p \leftarrow 2/n$ and return
      	to step~\ref{step:loop}.
\item \label{step:privamp}
	Both compute the number $\Delta$ of bits of information that Eve
      	could have obtained (taking into account bits exchanged in the
      	verification step(s)), perform privacy amplification as 
      	outlined in~\S\ref{sec:privamp}, and decrease $n$ accordingly.
\item \label{step:end}
      	If $n$ is sufficiently large, both consider the process successful,
      	otherwise reset $n$ (perhaps to a larger value than before)
      	and return to step~\ref{step:begin}.
\item \label{step:final}
	Both retain some of their $n$ bits for future use in authentication
	and as seeds for their random number generators; the remaining bits
	are available for use as a one-time pad or for other purposes.
\end{enumerate}

\subsection*{Notes}

The seed for the random number generator at step~\ref{step:seed} could be
derived from a previously shared key if this is not the first run (and
similarly for the random bits required for authentication on the classical
channel)~-- see step~\ref{step:final}. 
Note that Eve's chance of cracking the system is negligible unless
she can predict the random permutations that are used by Alice and Bob,
because without this knowledge the best she could obtain by eavesdropping on
both channels would be a random permutation of the final shared key.
\smallskip

In our simulations we found that a good strategy was to send a $64$-bit hash
with the parity bits at step~\ref{step:parities} whenever $p < 10/n$. If
their parities agree {\em and} the hashes agree, then Alice and Bob assume
that their reconciliation has been successful and proceed to
step~\ref{step:privamp}.

\section{Privacy Amplification}				\label{sec:privamp}

An important aspect of QKD is {\em privacy
amplification}, in which the block of bits that Alice and Bob have agreed on
is reduced in size to compensate for the information that Eve may have about
these bits.
\smallskip

More precisely, after Alice and Bob reach agreement on a block of say $m$
random bits, they need to estimate how many useful\footnote{We distinguish
between {\em useful} information, which is relevant to the bits that Alice
and Bob retain, and {\em useless} information, which is only relevant to
bits that Alice and Bob have discarded. We can assume that Eve's useful 
information per bit does not increase when Alice and Bob discard bad blocks
(in fact it is more likely to decrease, since eavesdropping tends to increase
Bob's error rate).} bits (say $\Delta$) of
information Eve could have gleaned, and reduce the size of their
agreed block by $\Delta$ bits using a process such as random subset
hashing\footnote{Random subset hashing is similar to the first hashing
method described in \S\ref{sec:verify}, with
$k = m-\Delta$.} (or give up if $m-\Delta$ is too small).  An upper bound on
$\Delta$ depends on the physics of the quantum communication and the
observed error rate.  For details see~\cite[Ch.~7]{Sharma}.
\smallskip

Conventional cryptography gives security by imposing a time-consuming
computational task on Eve.  Except in the case of the one-time-pad method,
Eve can break the system if she can perform enough computations to do a 
brute-force search through the key space. In practice, keys are chosen
large enough that this is impractical (at present).  However, it is difficult
to be confident that it will be impractical in the future. For example,
the RSA cryptosystem depends on the difficulty of factoring large integers,
but this has not been {\em proved} to be difficult. It is quite possible
that a practical polynomial-time algorithm for factoring exists (as it does
for the related problems of primality testing and factoring polynomials over
finite fields). Also, if a quantum computer can be built, then factoring
(and other problems of cryptographic interest such as discrete logarithm
problems) will be possible in polynomial time.
\smallskip

QKD, on the other hand, does not need to impose any limits on Eve's
computational power. It is only assumed that Eve has to obey the laws of
physics. By taking advantage of these laws and designing their system
correctly, Alice and Bob can detect any significant attempt by Eve to
eavesdrop on the quantum communication channel.
\smallskip

Alice and Bob still need to guard against a ``[wo]man-in-the-middle''
attack in which Eve intercepts their communications, 
impersonating Bob to Alice and Alice to Bob.  
For this reason, the classical channel between
Alice and Bob needs to be authenticated. This can be done using standard
techniques provided that Alice and Bob share an initial secret (of the 
order of a few hundred random bits). Using this secret for authentication,
Alice and Bob can ``bootstrap'' their system to generate a much longer shared
secret. This longer secret can be used as a one-time pad, or, if we are 
willing to trade off security against bandwidth, as
a key for a good stream cipher (see for example~\cite{Salsa}).
\smallskip

Since we assume that Eve has unbounded computational power, we should 
assume that Eve can break any encryption used on the classical channel
and eavesdrop on it successfully\footnote{This is not an argument for using
weak or no encryption on the classical channel. We should make life as 
difficult as possible for Eve by using strong encryption on the classical
channel. Even if Eve can crack this encryption, it should take her a
significant amount of time to do so, making it difficult for her to mount
a collective attack~\cite[\S12.6.5]{Nielsen-Chuang}.}.

\section{Bounding Eve's Information} 			\label{sec:eveinfo}

Before performing privacy amplification, Alice and Bob need to estimate (an
upper bound on) the amount of information (measured in bits) that Eve could
have obtained about their shared secret bit-string. Eve has two possible
sources of information\footnote{Apart from human error, physical theft,
etc.}~-- eavesdropping on the quantum channel, and eavesdropping on the
classical channel. As mentioned above, we assume that Eve can break any
encryption used on the classical channel. In particular, Eve can learn the
parities of blocks as they are exchanged by Alice and Bob
(step~\ref{step:parities} of the summary above). However, since she does not
know the seed for Alice and Bob's pseudo-random number generator, she can
not predict {\em in advance} the random permutations that Alice and Bob
apply\footnote{If she could predict these
permutations in advance, Eve could use this information to choose which bits
to eavesdrop on the quantum channel. Assume that the initial blocksize is
two, as in the example given in Table~\ref{tab:simulate}. 
Suppose that Eve learns one bit from each block of two bits (she can predict
which bits will be in each block from knowledge of the first permutation).
Then, once she learns the parities of the blocks, she can deduce the values
of all the bits that were transmitted over the quantum channel, even though
Alice and Bob might think that she only knows 50\% of them.}.
\smallskip

The physics of the quantum channel allows Alice and Bob to give an upper
bound on the number of bits $\Delta$ that Eve learns by eavesdropping on 
the quantum channel.  Let $p_e = \Delta/n$, so $p_e$ is the fraction of
bits that Eve knows (before parity information is exchanged).  
For example, in the setup of Bennett {\em et al}~\cite{Bennett92a},
$\Delta \le p\sqrt{8}$, where $p$ is the error rate observed by Alice
and Bob 
(this can be estimated as in \S\ref{sec:reest})\footnote{Here as elsewhere
we have ignored the fact that our estimate of Eve's knowledge is statistical
rather than deterministic. For safety we should include 
``five standard deviation'' terms. These have been omitted because they
are $O(n^{-1/2})$ and we assume that $n$ is large. However, such 
terms would need to be included in the final analysis.}.
\smallskip

The protocol used by Alice and Bob ensures that Alice's {\em relevant}
information $\Delta$ does not increase
as a result of Eve eavesdropping on the classical channel. For example,
whenever Eve learns the parity of a good block, one bit of that block is
discarded. If Eve did not already know that bit, her parity information
is useless. If she did know that bit, then she gains parity
information about the remaining bits in the block, but in compensation
she loses a bit of information about Alice and Bob's (retained) data.
In either case, her information (in the sense of Shannon's information 
theory) does not increase, although the actual information may change.
\smallskip

The fact that Eve's useful information does not increase is sufficient
for Alice and Bob's purposes
if $p_e$ and $p$ are sufficiently small. For example, consider
Table~\ref{tab:p25} or Table~\ref{tab:simulate}, which assume $p=0.25$
and $n = 10^6$.
If $p_e < 0.09$ then $\Delta \approx 90000$ but $n_f > 97000$ leaving an
adequate margin of at least $7000$ bits. Similarly, if $p = 0.1$ then we
expect $n_f > 310000$ so Alice and Bob can succeed even if
$p_e = p\sqrt{8} \approx 0.283$.
\smallskip

If $p_e$ is too large for this argument to be useful (for example, if $p_e
\ge 0.1$ with $p = 0.25$, see Table~\ref{tab:p25}), Alice and Bob can use a
different argument, which we now describe.  We consider two cases. In the
first case, which we assume occurs initially, Eve's information is about
individual bits. That is, Eve knows about $p_e n$ of the $n$ bits
transmitted from Alice to Bob.  Eventually (after Alice and Bob have
used a blocksize greater than two), Eve may have gained information 
in the form of
nontrivial linear relations (over $\GF(2)$) between bits by eavesdropping
on parity information that is exchanged on the classical channel. (Because
Alice and Bob discard a bit from each good block, Eve does not gain such
information while the blocksize is two.) 
If Eve gains enough such relations she can solve for
the unknown bits (or at least restrict a brute-force search to a
low-dimensional space) by performing linear algebra over $\GF(2)$.
Thus we have to count each linear relation as a bit of information. 
If Eve is expected to have $n_e$ bits of information about the $n$ bits 
that have not yet been discarded, then the current value of 
$p_e$ is $n_e/n$.  It is convenient to define $q_e = 1 - p_e$.
\smallskip
 
\subsection{Case 1: Eve knows only individual bits}

Consider the effect of a round with blocksize~$b$ in the first case (when
Eve knows some individual bits but no nontrivial relations).  
With probability
$q_e$, Eve does not know the first bit in a given block, so the parity
information in that block is useless to her (since the first bit will be
discarded). Thus, Eve's probability of knowing any of the remaining bits in
the block is unchanged. Also, with probability $p_e^b$, Eve already knows all
the bits in a given block, so the parity information tells her nothing new.
In the remaining cases, which occur with probability 
$1 - q_e - p_e^b = p_e - p_e^b$,
Eve already knows the first bit but not all bits in the block,
and she gains parity information about the remaining bits, that is a linear
relation satisfied by these bits. 
Thus, overall, the effect of one round is to replace $p_e$ by 
\begin{equation}
p_e' = p_e + \frac{p_e - p_e^b}{b-1}\;.			\label{eq:pecase1}
\end{equation}
Since
\[\frac{p_e - p_e^b}{b-1} = 
p_e q_e \left(\frac{1 + p_e + \cdots + p_e^{b-2}}{b-1}\right) \le p_e q_e\;,\]
we have $1 - p_e' = q_e' \ge q_e^2$. Equality holds iff
$b=2$ or $p_e = 0$ or $q_e = 0$.

\subsection{Case 2: Eve may know nontrivial relations}

Because a nontrivial relation involves two or more bits, the argument given
for Case~1 does not apply if Eve knows some nontrivial relations\footnote{It
is {\em plausible} that a nontrivial relation is no more use to Eve than
knowledge of a single bit, so~(\ref{eq:pecase1}) applies in all cases, 
but we can not prove this.}. 
In Case~2, Eve's knowledge might increase by one bit for
each parity block.  Thus,~(\ref{eq:pecase1}) has to be replaced by
\begin{equation}
p_e' = \min(1, p_e + 1/b)\;.			\label{eq:pecase2}
\end{equation}
Note that~(\ref{eq:pecase2}) applies whether or not Alice and Bob discard a bit
from each good block. However, it seems plausible that Eve's task is made
more difficult by such discards.

\subsection{Improved strategy for choosing the blocksize} \label{sec:block2}

The blocksize selection strategy considered in \S\ref{sec:blocksize}
may not work if $p_e$ is large (or equivalently, if $q_e$ is small). 
Note that no strategy can work if
$q_e \le p_e$, because this inequality can be interpreted as saying that
Eve's information is better than Bob's (and it will continue to be at least
as good if Eve can eavesdrop on the classical channel).  Thus, we have to 
assume that $q_e > p_e$. The strategy suggested below should work (in the 
sense of giving Alice and Bob a significant advantage over Eve) provided
there is some slack in this inequality.  Our simulations suggest that it
works if $q_e/p_e \ge 4$, and in some circumstances (depending on $p_e$ and
what we regard as a ``significant'' advantage) if $1 < q_e/p_e < 4$.
\smallskip

There are two (conflicting) requirements on the blocksize $b$. In order to 
reduce the error rate substantially each round (see equation~(\ref{eq:newp})),
Alice and Bob want to choose $b$ significantly smaller than $1/p$. 
On the other hand, in order not to give Eve
too much information in the form of parity bits, they want $b$ significantly
larger than $1/q_e$. Since we assume $p < q_e$, we have
$1/q_e < 1/p$, and we should choose $b \in (1/q_e, 1/p)$.
A reasonable compromise is to take the geometric mean,
that is $b = 1/\sqrt{p q_e}$.  Of course, we also have to restrict $b$
to be an integer (and at least two).
\smallskip

Simulations indicate that, if $q_e/p_e$ is close to $1$, it is best to
choose $b=2$ so that we stay in case~1 above and can use~(\ref{eq:pecase1})
instead of~(\ref{eq:pecase2}) to update the estimate $p_e$ of Eve's 
useful information per bit. 
While $b=2$, both $p$ and $q_e$ are 
approximately squared each round, so the ratio $q_e/p$ increases, although
both $p$ and $q_e$ decrease.
Once $q_e/p$ increases above some
threshold, it is possible to use a larger blocksize, even though this
means that case~2 applies in later rounds.  A good strategy is to
take
\begin{equation}
b = 
\begin{cases}
	2 		& \text{if case 1 (no relations) and $4p > q_e$,}\\
	\lfloor\max(2, 1/\sqrt{p q_e})\rfloor 
			& \text{otherwise.}
\end{cases}						\label{eq:bestb}
\end{equation}
Consider an example with $n = 10^6$, $p = 0.15$,
$p_e = 0.25$. The predicted outcome is shown in Table~\ref{tab:qc7k-tab6}.
The last column $(n' - \Delta')$ gives Alice and Bob's advantage over Eve. 
It can be seen that Alice and Bob end up with more than $88,000$ bits
(out of $211,767$ bits) that are unknown to Eve. Since Eve started with
knowledge of $250,000$ bits, using monotonicity of $\Delta$ would not
be sufficient.

\begin{table}
\centering
\caption{Prediction for $p=0.15$, $p_e=0.25$, $n=1000000$.} 
\label{tab:qc7k-tab6} 
 ~\\[0pt]
\begin{tabular}{|c|c|c|c|c|c|c|}
\hline
$p$  &    $b$ &    $n$   & errors & bad blks & $n'$ & $n' - \Delta'$ \\
\hline
0.150000 & 2   & 1000000 & 150000 & 127500 & 372505 & 198281 \\
0.030201 & 7   & 372500  & 11250  &  9405  & 262858 & 127321 \\
0.005721 & 18  & 262858  & 1504   &  1366  & 225031 & 97658 \\
0.000561 & 64  & 225031  & 126    &   122  & 213839 & 89576 \\
0.000020 & 347 & 213839  & 4     &      4  & 211767 & 88101 \\
\hline
\end{tabular}
\end{table}

Table~\ref{tab:qc7} shows the predicted advantage $n' - \Delta'$ 
for various $p$ and $p_e$, all for $n = 10^6$. 

\begin{table}
\centering
\caption{Predicted advantage for various $p$, $p_e$, $n = 1000000$.}
\label{tab:qc7}						    	
 ~\\[0pt]
\begin{tabular}{|c|c|c|c|c|}
\hline
$p_e \backslash p $  & 0.1 & 0.2 & 0.3 & 0.4 \\
\hline
0.0 & 247373 & 130017 & 56571 	& 13361 \\
0.1 & 203493 & 93049  & 31208 	& 3449  \\
0.2 & 158045 & 59548  & 8207   	& 217   \\
0.3 & 117032 & 34798  & 4492	& ---   \\
\hline
\end{tabular}
\end{table}

Table~\ref{tab:qc7e} shows the predicted advantage 
for various $p$ and the ratio $q_e/p \in \{2, 3, 4, 5\}$, also for $n = 10^6$. 
In the table, a dash means that the advantage is smaller than $64$.
It can be seen that the advantage is always significant if
$q_e/p \ge 4$, and can be significant even for $q_e = 2p$.

\begin{table}
\centering
\caption{Predicted advantage for various $p$ and $q_e/p$, $n = 1000000$.} 
\label{tab:qc7e}						
 ~\\[0pt]
\begin{tabular}{|l|c|c|c|c|}
\hline
$q_e\; \backslash\; p $  & 0.001 & 0.01 & 0.1 & 0.2\\
\hline
$2p$ & ---	&	---	& 94  	& 559 \\
$3p$ & --- 	&	109	& 6253 	& 15539 \\
$4p$ & 90	&	784	& 12139 & 59548 \\
$5p$ & 329	&	3237	& 40606	& 130017 \\
\hline

\end{tabular}
\end{table}

The number of bits communicated over the classical channel during the
verification phase(s) should be taken into account when estimating the 
information available to Eve. This would decrease the advantage predicted
in Tables~\ref{tab:qc7k-tab6}--\ref{tab:qc7e} by about $64$ bits (but the
change does not scale with $n$). 

\section*{Appendix A: Permutation Generators}

Alice and Bob should use a good pseudo-random permutation generator
such as the Durstenfeld shuffle.
This is often called the {\em Knuth shuffle}~\cite[Alg.~P]{Knuth}, 
but was first published by Durstenfeld~\cite{Durstenfeld}. It is
sometimes called the {\em Fisher-Yates shuffle}, but this is incorrect
because the algorithm
proposed by Fisher and Yates, while suitable for hand computation,
is inefficient on a computer~\cite{Fisher-Yates,wiki1}.
\smallskip

It turns out that, at least for large blocksizes, the most expensive part of
Alice and Bob's computation is performing random permutations.  This is
partly due to the fact that the permutation accesses bits at random
addresses in a ``cache-unfriendly'' manner. 
For the sake of efficiency we use
a ``cache-friendly'' permutation which
restricts the distance that bits may move to
less than a suitable fraction of the L2 cache size.  Since the L2 cache is
typically at least $64KB$, this is good enough, although the output is no 
longer uniformly distributed over all $n!$ possible permutations.

\end{document}